\documentclass[twocolumn,showkeys,preprintnumbers,amsmath,amssymb]{revtex4}

\usepackage{graphicx}
\usepackage{dcolumn}
\usepackage{bm}

\begin{document}

\preprint{APS/123-QED}

\title{First-principles study of the layered thermoelectric material TiNBr}

\author{Shuofeng Zhang}
\author{Ben Xu}%
 \email{xuben@mail.tsinghua.edu.cn}
 \altaffiliation{Key Laboratory of Advanced Materials of Ministry of Education of China, People's Republic of China}
\author{Yuanhua Lin}
 \altaffiliation{State Key Laboratory of New Ceramics and Fine Processing Beijing 100084, People's Republic of China}
\author{Cewen Nan}
 \altaffiliation{State Key Laboratory of New Ceramics and Fine Processing Beijing 100084, People's Republic of China}
\author{Wei Liu}
 \altaffiliation{Key Laboratory of Advanced Materials of Ministry of Education of China, People's Republic of China}
\affiliation{%
School of Materials Science and Engineering \\
Tsinghua University, People's Republic of China
}%


\date{\today}

\begin{abstract}
Layer-structured materials are often considered to be good candidates for thermoelectric materials,
because they tend to exhibit intrinsically low thermal conductivity as a result of
atomic interlayer interactions. The electrical properties of layer-structured materials can be easily tuned using various methods,
such as band modification and intercalation.
We report TiNBr, as a member of the layer-structured metal nitride halide system MNX (M = Ti, Zr, Hf; X = Cl, Br, I),
and it exhibits an ultrahigh Seebeck coefficient of 2215 $\mu V/K$ at 300K.
The value of the dimensionless figure of merit, ZT, along the A axis can be as high as 0.661 at 800K,
corresponding to a lattice thermal conductivity as low as 1.34 W/(m K). The low ${\kappa_l}$ of TiNBr is associated
with a collectively low phonon
group velocity ($2.05\times 10^3 $ m/s on average) and large phonon anharmonicity
that can be  quantified using the Gr\"{u}neisen parameter and three-phonon processes.
Animation of the atomic motion in highly anharmonic modes mainly involves the motion of N atoms,
and the charge density difference reveals that the N atoms become polarized
with the merging of anharmonicity. Moreover, the fitting procedure of the energy-displacement curve verifies
that in addition to the three-phonon processes, the fourth-order anharmonic
effect is also important in the integral anharmonicity of TiNBr. Our work is the first study of the thermoelectric properties of TiNBr
and may help establish a connection between the low lattice thermal conductivity and the behavior of phonon vibrational modes.
\end{abstract}

\maketitle

\section{\label{sec:level1}Introduction\protect
}

TiNBr belongs to a series of layer-structured metal nitride halides of the form MNX, where M is Ti, Zr, or Hf and X is Cl, Br, or I. Many researchers have systematically investigated the superconductivity and the intercalation situations of these systems \cite{RN9,RN10,RN60,RN61,RN62,RN63,RN64}.
This kind of compound has easily tunable electrical properties, and their layered structure may help them introduce extra phonon scattering processes via atomic interlayer interactions, which tend to cause the materials to exhibit intrinsically low thermal conductivity \cite{RN6, RN53, RN54}, thus making them possible candidates as thermoelectric materials.

\par
Thermoelectric materials enable direct conversion between thermal and electrical energy, and can generate environmentally-friendly energy from waste heat. Normally, we use the dimensionless figure of merit
($ZT$) to characterize the thermoelectric conversion efficiency. The $ZT$ value is defined as $ZT = (S^2\sigma/\kappa)T$,
 where $S$, $\sigma$, $\kappa$ and $T$ are the Seebeck coefficient, electrical conductivity, thermal conductivity, and
absolute temperature, respectively. In the fractional term of $ZT$, the numerator $S^2\sigma$ is also called the power factor.
To improve the value of $ZT$, its simple expression provides researchers with explicit objectives, such as  increasing the power factor
and decreasing the thermal conductivity \cite{RN55}. However, the well-known interdependence between $S$, $\sigma$ and $\kappa$
complicates the procedure. Consequently, we often try to maintain a high power factor and to reduce the lattice thermal
electricity, or to optimize the power factor in materials that have an intrinsically low thermal conductivity \cite{RN11}.\par
In recent decades, a large number of thermoelectric materials have been discovered. A common feature among
many of the high performance thermoelectric materials, such as BiCuSeO, is that many of them have a layered
crystal structure.
As mentioned above, TiNBr is a kind of material that
has a layered and anisotropic crystal structure \cite{RN31, RN9, RN10}. Because the thermoelectric properties of TiNBr have not been systemically investigated, we conducted first-principle calculations on the electronic transport properties and thermal transport properties of TiNBr, and surprisingly found that it exhibits attractive thermoelectric properties.
 The $ZT$ value of TiNBr along A axis can be as high as 0.661 at 800K when the electronic relaxation time was conservatively set
  as 0.8$\times 10^{-14}$ s, and the corresponding lattice thermal conductivity can be low as 1.34 W/(m K).
 \par
 It is known that phonons play an important role in heat transfer and influence the tuning of the thermal conductivity. For instance, the leading thermoelectric material, PbTe, exhibits giant anharmonic phonon scattering
 \cite{RN56}, and resonant bonding leads to low thermal conductivity in rock salt IV-VI compounds such as SnTe and
Bi$_2$Te$_3$ \cite{RN57}.
 To understand the reason behind the intrinsically low lattice thermal conductivity of TiNBr from a theoretical
 perspective, we additionally explored the relationship between the phonon behavior and the thermal conductivity based
 on the mode level phonon group velocity and Gr\"{u}neisen parameter analysis.
 The
 anharmonic phonon scattering of different vibration modes was also investigated \cite{RN2, RN3}.
 The data indicates that TiNBr is a prospective thermoelectric material, that consists of
 Earth-abundant and harmless elements.

\section{Computational details}
For the calculations, we used Density Functional Theory (DFT) based on first principles as implemented in the Vienna ab initio Simulation Package (VASP) \cite{RN18, RN19}. 
The Local Density Approximation (LDA) + U using the projected augmented wave (PAW) method was used for the exchange and correlation effects \cite{RN26, RN27}.
The LDA+U method takes into account orbital dependence of the Coulomb and exchange interactions which is absent in the LDA.
However, LDA+U still tends to underestimate band gaps because it insufficiently accounts for exchange-correlation effects of the localized 3d electrons, and so we also adopted the HSE06 hybrid functional to calculate the band structure. The HSE06 hybrid functional is based on a screened Coulomb potential for the exchange interaction \cite{RN7,RN8}.
$E_{xc}$ is given by
\begin{eqnarray}
E_{xc}^{\omega PBEh} = aE_x^{HF,SR}(\omega)+(1-a)E_x^{PBE,SR}(\omega)\nonumber\\+E_x^{PBE,LR}(\omega)+E_c^{PBE}
\end{eqnarray}
where $\omega$ is an adjustable parameter governing the extent of short-range interactions. The $\omega$PBE
hybrid ($\omega$PBEh) is equivalent to PBE0 for $\omega = 0$ and asymptotically reaches PBE for $\omega\rightarrow \infty$.
\par
The wave functions are expanded in a plane wave basis with a kinetic energy cut off of 500 eV.
A Monkhorst-Pack k-mesh of $39 \times 39 \times 19$ was used to sample the Brillouin Zone (BZ),
 and the energy and atomic force convergence thresholds were set as $10^{-6}$ eV and $10^{-3}$ eV\AA$^{-1}$, respectively \cite{RN32}.
\par
The BoltzTrap code \cite{RN20} was used to calculate the electronic transport properties.
Using this code, the thermoelectric properties, including the electrical conductivity ($\sigma$) and electronic
thermal conductivity ($\kappa_e$), were calculated with respect to the relaxation time ($\tau$), which means that we
get $\sigma/\tau$ and $\kappa_e/\tau$ from BoltzTrap. The $ZT$ value can be calculated as:
\begin{eqnarray}
ZT = \frac{S^2\sigma }{\kappa}T = \frac{S^2\frac\sigma \tau \cdot \tau}{\kappa_l + \frac{\kappa_e}{\tau}\cdot\tau}T
= \frac{S^2\frac\sigma \tau }{\frac{\kappa_l}{\tau} + \frac{\kappa_e}{\tau}}T
\end{eqnarray}
 If $\sigma/\tau$ and $\kappa_e/\tau$ are fixed, the final value of $ZT$  increases when the relaxation time increases.
We adopted a conservative value of $\tau = 0.8\times 10^{-14}$ s to calculate the $ZT$ value, and this also ensures that the result tends to be underestimated. \par

The harmonic and anharmonic interatomic force constants (IFCs) were obtained using the real-space finite displacement
difference method.
A $3\times 3\times2$ supercell that contained 108 atoms was constructed for the force calculations, and the  energy and atom force convergence thresholds were set as
$10^{-8}$ eV and $10^{-8}$ eV\AA$^{-1}$, respectively.
A Monkhorst-Pack k-mesh of $6\times6\times4$ was used to sample the BZ
using the Phonopy package \cite{RN21}.
Space group symmetry properties were used to reduce the calculation cost and numerical noise of the force constants; they can also greatly simplify the determination of the dynamical matrix constructed on the basis of the harmonic IFCs \cite{RN46, RN47}.


\par
The calculation of lattice thermal
conductivity was conducted using the ShengBTE code \cite{RN12} based on the phonon Boltzmann transport equation (pBTE).
This code uses the second-order (harmonic) and third-order (anharmonic) IFCs
combined with a full solution of the pBTE to successfully predict the lattice thermal conductivity ($\kappa_l$) \cite{RN33, RN44}.
A Monkhorst-Pack k-mesh of $2\times2\times1$ was used to sample the BZ, and
a cutoff radius $r_{cutoff}$ was introduced to
disregard interactions between atoms that had a distance larger than a certain value for practical purposes \cite{RN38, RN47}.
Based on the convergence test of $\kappa$ with varying values of $r_{cutoff}$,
we chose the cutoff interactions to include up to the $8^{th}$ nearest neighbors
(corresponding to the cutoff distance of $ 4.6864$ \AA) and a Q-grid of $ 11 \times 11\times 5$ for calculating the $\kappa_l$
of TiNBr. Additionally, the Born effective charges ($Z^*$) and dielectric constants ($\epsilon$) were obtained using
the density functional perturbation theory (DFPT), which is added to the dynamical matrix as a correction to account for long-range electrostatic interactions.

\section{Results}
\subsection{Structural information}
\begin{figure}
\includegraphics{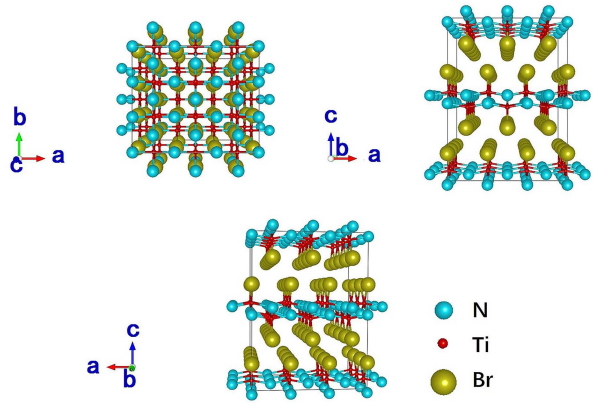}
\caption{\label{fig:1} Crystal structure of TiNBr. The superlattice shows a $3\times3\times2$ primitive cells. Representations of the different atoms are noted in the legend.}
\end{figure}

TiNBr materials crystallize in the P\emph{mmn} space group. The optimized geometry structure is shown in Fig.
\ref{fig:1}.
The calculations were performed on the primitive cell containing six atoms (two atoms for each element).
The positions of the N atoms are (0, 0, 0.950) and (0.5, 0.5, 0.050), those of the Ti atoms are (0.5, 0, 0.904) and (0, 0.5, 0.096), and those of the Br atoms are (0, 0, 0.338) and (0.5, 0.5, 0.662).
Several different methods were employed to optimize the structure including PBE and LDA, both with or without +U, and we found LDA+U can achieve the best result compared with experiment. According to research by Chang-Hong Chien \emph{et al}, The Coulomb parameter, U, and exchange terms (screened), J, for the Ti atoms were chosen as
6.6 eV and 0.78 eV, respectively\cite{UJ}.
The calculated value of the lattice constants compared to the experimental results reported by Zhang \emph{et al.}
is given in Tab. \ref{tab:table1}\cite{RN10}.
The optimized lattice constants of TiNBr are in good agreement with the experimental results, and this confirms the reliability of the method used.

\begin{table}
\caption{\label{tab:table1}Calculated and experimental lattice constants of TiNBr.  }
\begin{ruledtabular}
\begin{tabular}{lccr}
 & $a$(\AA)  & $b$(\AA)&$c$(\AA)\\
\hline
This work  & 3.999 & 3.359 & 8.139  \\
Experimental  & 3.927 & 3.349 & 8.332 \\
\end{tabular}
\end{ruledtabular}
\end{table}


\subsection{Electrical properties}
\begin{figure*}
\includegraphics{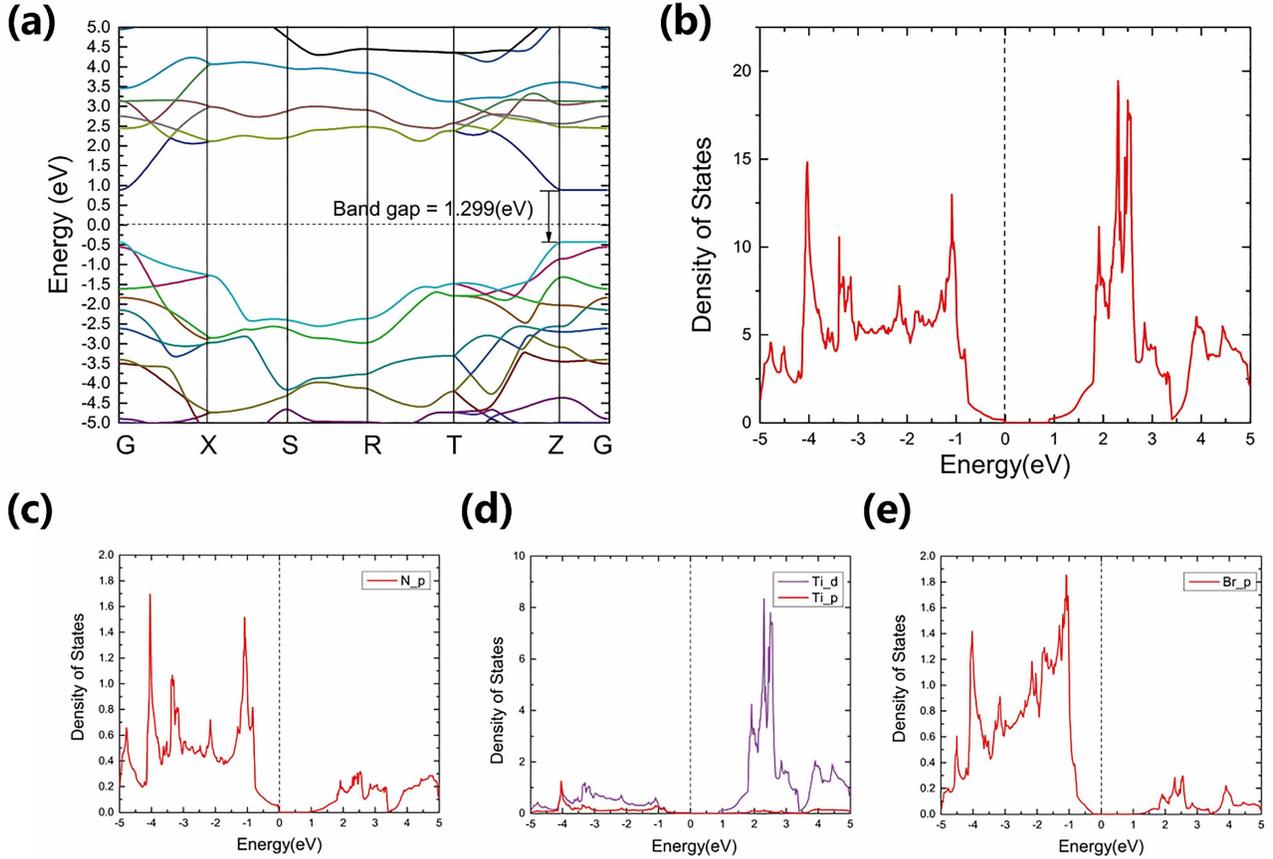}
\caption{\label{fig:2}Calculated band structure (a) and (projected) density of states near the Fermi energy (b)-(e) for TiNBr. The dashed lines represent the Fermi levels and are set to zero.}
\end{figure*}

The calculated electronic band structure of TiNBr and the density of states (DOS) are shown in Fig. \ref{fig:2}.

TiNBr is a direct-band gap semiconductor that has both the conduction band minimum (CBM) and the
valence band maximum (VBM) at point Z. Using HSE06, the band gap is calculated as 1.299 eV, and this is much larger than the LDA+U value of 0.883 eV. Because of the superiority
of the HSE06 hybrid functional in calculating band gaps and because of the lack of an experimental value, we use the HSE06 result
as input in the subsequent calculations if a band gap value is necessary.
\par

TiNBr tends to form p-type semiconductors. Thus, the upper
part of the valence band has the greatest effect on the electrical properties \cite{RN6}.
As shown in Fig. \ref{fig:2}(a), the overall dispersion has relatively low curvature, especially along the Z - $\Gamma$
line, where the curves are almost straight and flat, indicating a large carrier effective mass around the VBM \cite{RN4}.
Previous studies have proven that the light band provides good electrical conduction, and
the heavy band is favorable for a high Seebeck coefficient \cite{RN11, RN51, RN52}.

\par

The total DOS near the band gap and the projected density of states (PDOS) are shown in Figs. \ref{fig:2}(b)-(e).
Although rapid changes in the DOS represent
a large Seebeck coefficient \cite{RN11, RN28}, one can infer from FIG. 3(a) that the corresponding carrier density is rather small around the Fermi level,
and this results in a moderate power factor. Conducting
non-equivalent doping may be a
promising approach for optimizing the electrical conductivity and for obtaining good thermoelectric properties.
The PDOS give the detailed contribution of the atomic orbitals that compose in TiNBr.
The s-orbitals of Ti, N and Br make a negligible contribution.
The p-orbitals of N and Br contribute more in the valence band while Ti-d orbitals dominate in the conduction band.
This information provides a lot of inspiration if we want to modify the electrical properties of TiNBr.
\par
In addition, because TiNBr has a layered structure, intercalation may also help promote the electrical properties
without introducing too many negative effects with respect to the thermal properties. Zhang \emph{et al.} reported that, when
TiNBr was electron-doped by the intercalation of alkali metal, it can transform into a superconductor \cite{RN10}.
Because of the excellent electrical conductivity of copper, we tried to intercalate Cu atoms into the interlayer gap between the two adjacent bromide layers in a proportion of TiNBrCu$_{0.25}$, but the concentration of copper was too high to maintain the semiconductive character of the system, and the thermal conductivity was consequently also so high that the whole ZT value was not improved. Although we did not study this concept in more depth, reducing the concentration of the intercalation atoms or choosing other elements that have electronegativity more similar to bromine may help improve the performance.

\par
\begin{figure}
\includegraphics{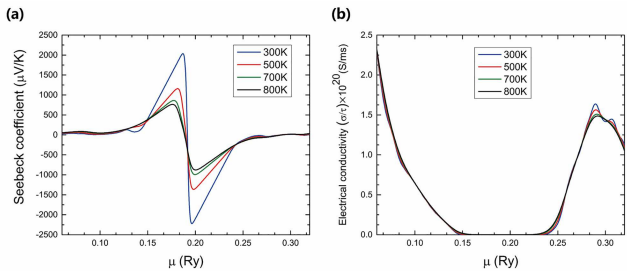}
\caption{\label{fig:3} The variation of (a) the Seebeck coefficient, (b) the electrical conductivity with respect to chemical potential ($\mu$) at different temperatures.}
\end{figure}
At different temperatures, the electrical properties of TiNBr as a function of chemical potential are shown in  Fig. \ref{fig:3}. A positive value of the chemical potential ($\mu$) compared to the Fermi level
indicates n-type doping, and a negative value of $\mu$ indicates p-type doping.
The Seebeck coefficient as a function of chemical potential at different temperatures (300, 500, 700, and 800K)
is shown in Fig. \ref{fig:3}(a).
The absolute value of the maximum Seebeck coefficient value is very high (2215 $\mu V/K$) and is obtained near the Fermi level at 300K, corresponding to n-type doping.
In Fig. \ref{fig:3}(a), the peaks of the Seebeck coefficient curves at different temperatures are near the Fermi level; at such points, the Seebeck coefficient dominates the overall electrical properties of a semiconductor. It is also obvious that the maximum value of the Seebeck coefficient decreases with an increase in temperature.

\par

Fig. \ref{fig:3}(b) shows the variation of the electrical conductivity with chemical potential and temperature. The electrical conductivity increases slightly as the temperature goes up because of the increase in the carrier density. The value of the electrical conductivity is higher for the negative chemical potential compared to the positive one, indicating that the p-type composition has higher electrical conductivity than the n-type doping has. The overall impact of temperature on the electrical conductivity is not comparable to that of the chemical potential.

\subsection{Phonon dispersion and thermoelectric properties}
\begin{figure}
\includegraphics{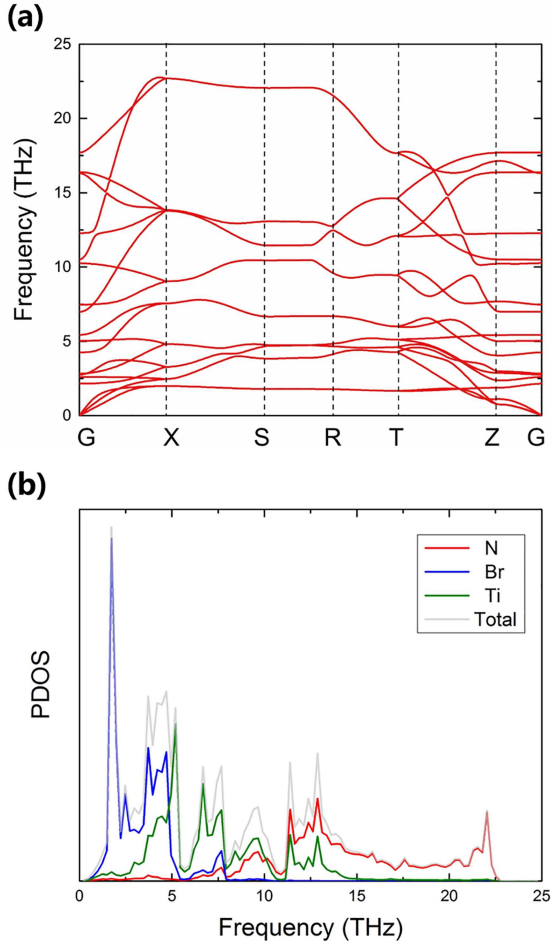}
\caption{\label{fig:4} The (a) phonon spectrum and corresponding (b) PDOS of TiNBr.}
\end{figure}

To investigate the stability and thermal properties of TiNBr, we first calculated the phonon spectrum using a $3\times3\times2$ superlattice, as shown in Fig. \ref{fig:4}(a). There are no imaginary frequencies, which proves that this particular structure is thermodynamically stable. The primitive cell of TiNBr contains 6 atoms. Thus, the calculated phonon spectrum has 18 dispersion relations. The phonon spectrum reveals the lattice vibration modes of TiNBr. Analyzing the atomic vibrations provided by Phonopy, the motion of each atom can be identified , and the interactions between atoms can be depicted in detail. Normally, the upper branches contribute more from lighter atoms, and the lower branches contribute more from heavier atoms. This theory is also applicable in the present study, as shown in Fig. \ref{fig:4}(b). Specifically, the N atoms dominate in Fig. \ref{fig:4}(b) upper branches where the frequency is higher than 10 THz, and the Br atoms mainly contribute to the lowest branches where the frequency is lower than 5 THz. The middle weight Ti atoms dominate the middle frequency region. Additionally, the heights of the peaks in Fig. \ref{fig:4}(b) also indicate the density of degeneracy, and this accounts for the intensity of the interactions between the different branches. The total DOS is consistent with the phonon spectrum. Moreover, the phonon spectrum has relatively flat curves, indicating a small phonon group velocity based on the formula $V_g = d\omega/dq$; this is beneficial for the low thermal conductivity. A visible explanation regarding the phonon group velocity in TiNBr is provided later.

\par

\begin{figure}
\includegraphics{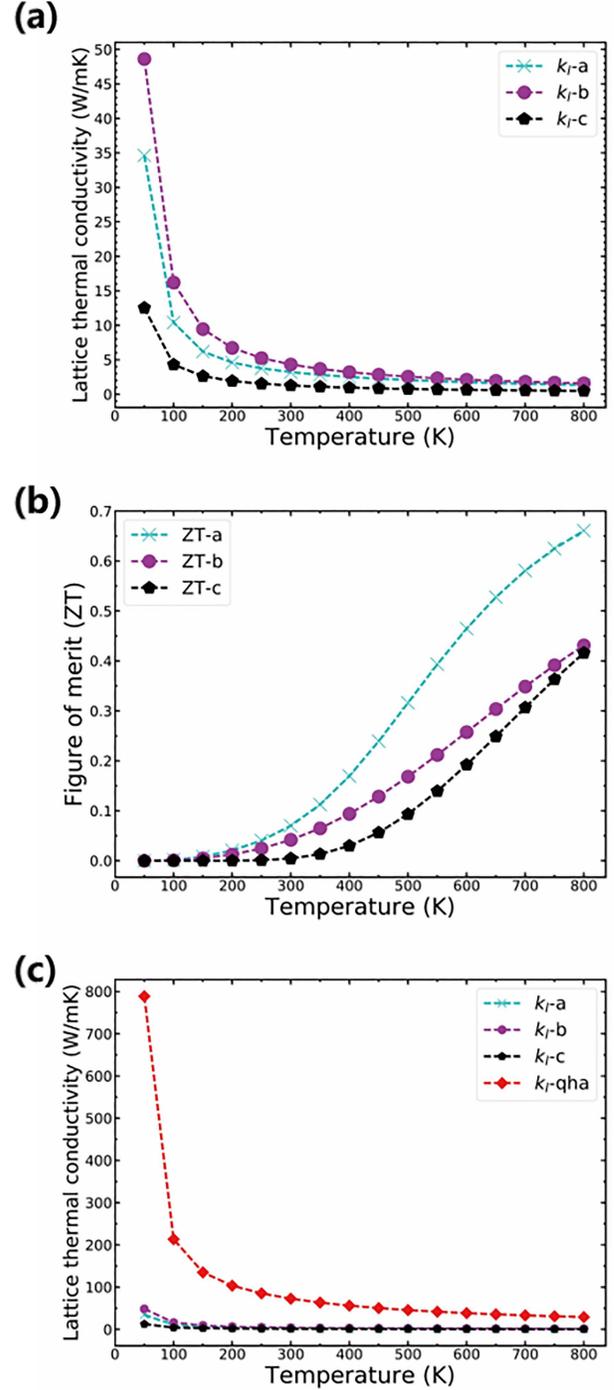}
\caption{\label{fig:5} The values of the lattice thermal conductivity and the values of ZT for TiNBr along the a, b, and c axes. (a) The lattice thermal conductivity along the different axes of TiNBr. (b) The values of ZT along the different axis of TiNBr. (c) Comparison of the calculated lattice thermal conductivity using the quasi-harmonic approximation and an approach that accounts for nonharmonic effects.}
\end{figure}

The anisotropy lattice thermal conductivities ($\kappa_l$) of TiNBr are provided in Fig. \ref{fig:5}(a). In general, the total thermal conductivity is the sum of the electronic thermal conductivity, $\kappa_e$, and the lattice thermal conductivity, $\kappa_e$. Because the electronic thermal conductivity of TiNBr is orders of magnitude smaller than the lattice thermal conductivities, only the latter one makes a significant contribution to the integral thermal conductivity. As shown in Fig. \ref{fig:5}(a), the value of $\kappa_l$ for TiNBr converges quickly when the temperature is higher than 300K. The values of $\kappa_l$ for the a, b, and c axes at 800K are 1.34 W/mK, 1.59 W/mK, and 0.49 W/mK, respectively. Although the value of $\kappa_l$ along the c axis is the lowest of the three, the electrical conductivity along this direction is too low to  ensure good thermoelectric performance. We care more about the intrinsic low lattice thermal conductivity along the other two directions, which are attributed to the small phonon group velocity and strong phonon-phonon scattering.

\par
With the combination of the electrical and thermal properties, the major criterion for the performance of a  thermoelectric material, namely the figure of merit (ZT), can be calculated. Fig. \ref{fig:5}(b) shows how the ZT varies with temperature.
The chemical potential is 0.23916 Ry at which the ZT value is highest, and the carrier concentration is 0.0313 per unit cell, which corresponds to a moderate Seebeck coefficient and to electrical conductivity with n-doping. The ZT values along the a, b, and c axes at 800K are 0.66, 0.43, and 0.42, respectively. The Seebeck coefficient generally decreases with an increase in temperature because of the emergence of more thermally excited charge carriers. However, at the particular chosen chemical potential, the decreasing rate at which the Seebeck coefficient changes with temperature is much slower than that of the lattice thermal conductivity. Thus, the overall ZT value gradually increases with an increase in temperature. The calculated ZT value is based on a conservative relaxation time constant that we chose as 0.8$\times10^{-14}$s. When the temperature increases, this approximation may be less effective because relaxation time decreases remarkably at high temperature. Thus, we set the upper limit of temperature at 800K, and estimations beyond this temperature are less significant.

\par

In fact, in Fig. \ref{fig:5}(a), the method we use to calculate the lattice thermal conductivity takes the nonharmonic effects into account to get a more reasonable result. For comparison, we also conducted the same calculation using the quasi-harmonic approximation (QHA). In this approximation, the lattice thermal conductivity can be estimated using the following relationship \cite{RN13}.
\begin{equation}
\kappa_l = \frac{A\overline{M}\theta^3\delta}{\gamma^2n^{\frac23}T}
\end{equation}
where $n$ is the number of atoms in the primitive cell, $\delta^3$ is the volume per atom, $\theta$ is the Debye temperature, $\overline{M}$ is the average mass of the atoms in the crystal, and $A$ is a collection of physical constants. (Specifically, $A\approx 3.1\times 10^{-6}$ if $\kappa$ is in W/mK, $\overline{M}$
in atomic mass unit, and $\delta$ in \AA). Here $\gamma$ is the temperature-dependent average Gr\"{u}neisen parameter, and it can be obtained from the QHA; a visualization of the data is given in the next section. The comparative results are shown in Fig. \ref{fig:5}(c), in which the results of the QHA calculations are depicted in red symbols. It is clear that the QHA calculated lattice thermal conductivity is far greater than that calculated using the nonharmonic effects, and this indicates that intrinsic scattering effects are not included in the QHA. This effect is favorable for the intrinsically low thermal conductivity of TiNBr, which is discussed in detail in the next section.

\subsection{Mode level analysis for the low {\bm{${\kappa_l}$}}}
\begin{figure}
\includegraphics{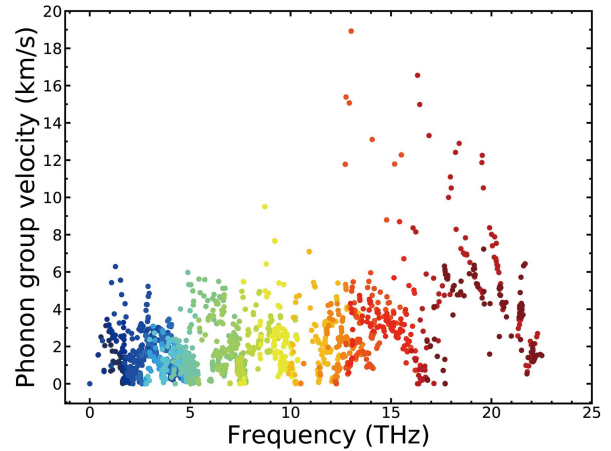}
\caption{\label{fig:6} Mode level phonon group velocity $v_g$ of TiNBr.}
\end{figure}

To explain the low lattice thermal conductivity, we perform a detailed analysis based on the mode level phonon group velocity, Gr\"{u}neisen parameter, phonon scattering rate, and vibrational mode. The mode level phonon group velocity is shown in Fig. \ref{fig:6}, and the different colors of the scattering data points correspond to different phonon modes or to different branches of the phonon spectrum. Colors near blue are used to depict the modes that have lower frequency, and colors near red represent the modes that have higher frequency. Thus, we can see intuitively the contributions of different modes and make comparisons between them. The same coloring scheme for the data points was also adopted for the following figures. As Fig. \ref{fig:6} clearly shows, most data correspond to a phonon group velocity lower than $6\times 10^3$ m/s. It is only in the high frequency region that some data points appear with high phonon group velocity. The average phonon group velocity in the irreducible wedge is only $2.05\times 10^3$ m/s, and this means that the phonon is generally transported slowly in the TiNBr lattice irrespective of the vibrational mode and position in Q-space. This partly contributes to low lattice thermal conductivity. The flattening of the phonon dispersion curves, as shown in Fig. \ref{fig:4}(a), is part of the reason behind the collectively low phonon group velocity.

\par

\begin{figure}
\includegraphics{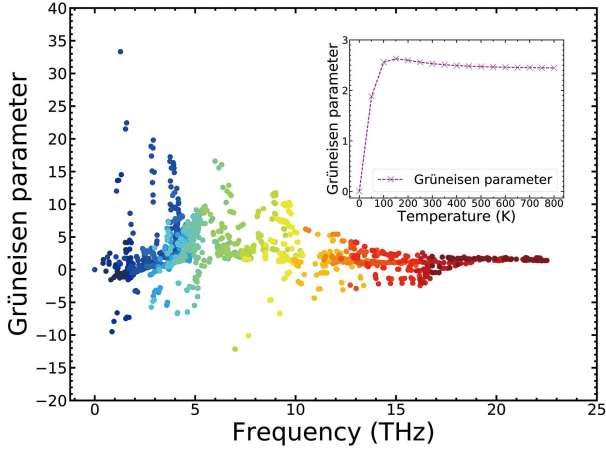}
\caption{\label{fig:7} Gr\"{u}neisen parameter of TiNBr. Inset: the temperature-dependent average Gr\"{u}neisen parameter of TiNBr.}
\end{figure}

For further insight regarding the atomic anharmonic interactions, phonon-phonon scattering and the Gr\"{u}neisen parameter ($\gamma$) were considered. It is well known that the phonon-phonon scattering process is determined by the anharmonic nature of structures \cite{RN42}, and the Gr\"{u}neisen parameter ($\gamma$) is often considered to be an anharmonicity parameter that reflects how much the phonon vibrations in a crystal lattice deviate from harmonic oscillations. Therefore, the magnitude of the anharmonicity in the TiNBr lattice can be roughly quantified by the Gr\"{u}neisen parameter. To this end, we examined the phonon anharmonicity of TiNBr by calculating the Gr\"{u}neisen parameter, and the results are shown in Fig. \ref{fig:7}. The temperature-dependent average Gr\"{u}neisen parameter of TiNBr is also plotted and shown in the inset figure. The value of $\gamma$ tends to be high when the frequency is low, indicating relatively strong anharmonicity because of the strong interactions between the acoustic and optical branches, as discussed earlier. The value of $\gamma$ decreases with an increase in frequency, indicating that the interaction between different vibrational modes becomes milder. For the temperature-dependent average Gr\"{u}neisen parameter, one can observe that the Gr\"{u}neisen parameter tends to increase sharply in the low temperature region, and gradually decrease to a stable value of 2.45 at 800K.

\par

\begin{figure}
\includegraphics{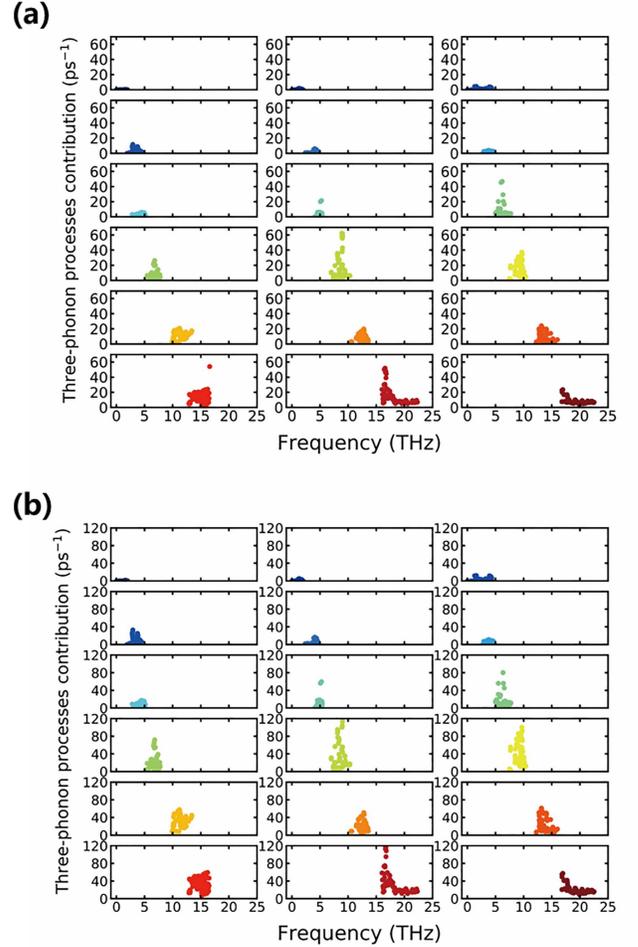}
\caption{\label{fig:8} Contributions of mode level three-phonon processes to the scattering rate at (a) 300K and (b) 800K. Each subfigure corresponds sequentially to a certain branch of the phonon vibrational mode.}
\end{figure}

Furthermore, we delve into the detailed processes of phonon-phonon scattering to gain a more fundamental understanding of the mechanisms underlying the low $\kappa_l$. Normally, of the kinds of anharmonic schemes that contribute to the final scattering rate, three-phonon processes play a vital role and are much more important than higher-order effects \cite{RN15,RN57,RN59}. Therefore, understanding the three-phonon induced scattering process can provide insight regarding anharmonicity. Three-phonon process contributions to the scattering rate with respect to different vibrational modes are shown in Fig. \ref{fig:8}, and each subfigure corresponds sequentially to a certain branch. The longitudinal scale plates are all set to the same value, and then the contributions of different branches can be compared in a straightforward way. Fig. \ref{fig:8}(a) shows the situation where the temperature is 300K, and Fig. \ref{fig:8}(b) shows that of 800K. In Fig. \ref{fig:8}(a), we can see clearly that of all of the vibrational modes, the 11$^{th}$ vibrational mode (second of the fourth row) has the maximum three-phonon process contribution to the scattering rate. The Ti, Br, and N atoms all join the lattice vibration, and the Ti atoms have the largest contribution, as seen in Fig. \ref{fig:4}. This particular kind of collective vibration has the largest incidence of three-phonon processes and corresponds to large anharmonicity. As surprising as it seems, the 17$^{th}$ vibrational mode also has a high three-phonon process contribution to the scattering rate, and only N atoms dominate the vibration. The case at 800K that is depicted in Fig. \ref{fig:8}(b) shows little change compared to that at 300K, except that the integral scattering rate increases because of the thermal excitation, and several new modes, such as the 12$^{th}$ and 17$^{th}$ modes, can have large enough contributions to be roughly the same as the 11$^{th}$ vibrational mode. Indeed, in comparing the scattering rates at different temperatures, the difference generally decreases with an increase in frequency, and this is also consistent with the trend in the average Gr\"{u}neisen parameter.

\par

\begin{figure}
\includegraphics{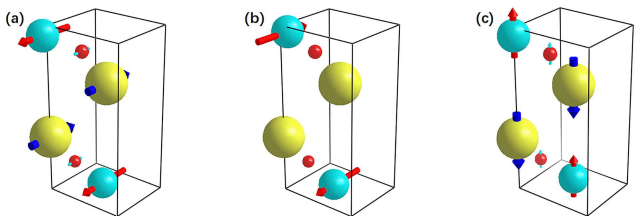}
\caption{\label{fig:9} Atomic motions of Ti, N, and Br atoms are shown in red, blue, and yellow, respectively. The arrows indicate the initial vibration directions, and the lengths of the arrows indicate the relative amplitudes. (a) 11$^{th}$ vibrational mode, (b) 17$^{th}$ vibrational mode, and (c) 12$^{th}$ interlayer vibrational mode.}
\end{figure}

To gain more detailed information regarding the vibrational mode, we additionally observed the vibrational motions of the atoms in the 11$^{th}$ and 17$^{th}$ vibrational modes using animation. Screenshots from these animations are shown in Fig. \ref{fig:9}. The atomic motions of the Ti, N, and Br atoms are shown in red, blue, and yellow, respectively. The arrows indicate the initial vibrational directions, and the lengths of arrows indicate the relative amplitudes. It is surprising to observe that these modes, which all have relatively large three-phonon process contributions, all mainly involve motion of the N atom. This is indicated by the comparatively long arrows that represent the motion of the N atoms. Indeed, in the animation, the N atoms move in the same phase or converse phases for the 11$^{th}$ and 17$^{th}$ vibrational modes, respectively, while the other atoms barely move. This means that the high three-phonon process contribution is mainly because of the motion of the N atoms.

\par

The 12$^{th}$ mode is another special mode, and it exhibits interlayer atomic motion as depicted in Fig. \ref{fig:9}(c). As shown, the Ti and N atoms form a layer and vibrate at the same pace along the c axis, and the Br atoms form another layer and occupy a converse phase. The influence interlayer atomic motion has on vibrational mode is the advantage of a layered material, and this particular mode contributes an additional scattering rate to the phonons. Fig. \ref{fig:8}(a) indicates that this kind of phonon scattering process has a pretty large contribution, ranking fourth among all of the vibrational modes at 300K, and when the temperature increases to 800K, the improvement to the phonon scattering rate induced by this interlayer atomic motion is more significant than that by any other modes. Thus, the interlayer effect is important at high temperature.

\par

\begin{figure}
\includegraphics{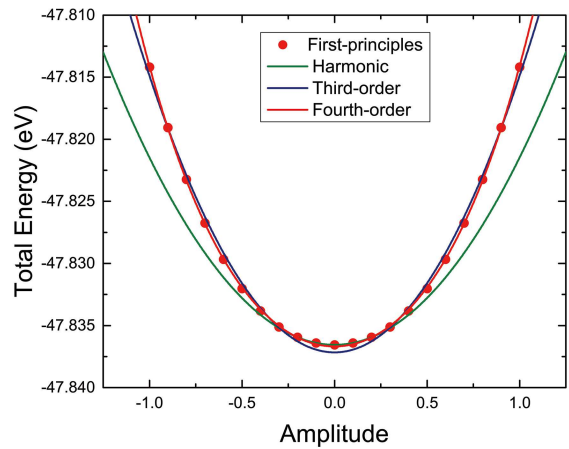}
\caption{\label{fig:10} Analysis of phonon anharmonicity in TiNBr using an energy-displacement curve, namely the total energy (together with harmonic and higher order fittings) with respect to the amplitude of a particular vibrational mode.}
\end{figure}

\begin{figure}
\includegraphics{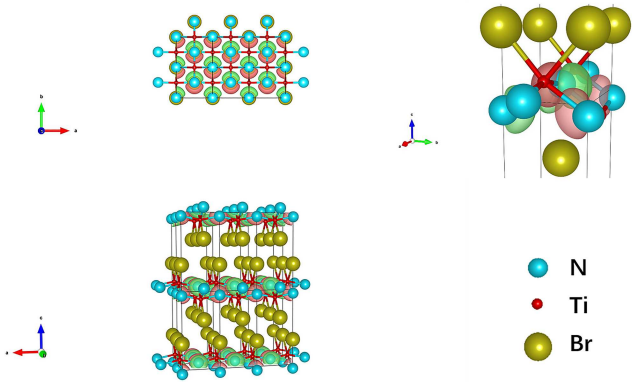}
\caption{\label{fig:11} Charge density differences between the initial position and the position where the energy-displacement curve starts to deviate from the harmonic curve. Red and green parts correspond to positive and negative charge accumulation, respectively.}
\end{figure}

The harmonic or anharmonic components of the energy-displacement curve can also be used to determine the origin of the anharmonicity in the TiNBr lattice. Specifically, here we use the amplitude of a particular vibrational mode (the 11$^{th}$ vibrational mode because of its maximum three-phonon process contribution) as the criterion for atom displacement to guarantee consistency with the corresponding phonon mode. The total energy computed using first-principles varies with the change in amplitude as shown in Fig. \ref{fig:10}. We use polynomial functions of different orders to fit the curve. The original harmonic function does not fit the curve, indicating the anharmonicity of TiNBr \cite{RN53, RN58}. The charge density difference between the initial position and the position where the energy-displacement curve starts to deviate from the harmonic curve is investigated and depicted in Fig. \ref{fig:11}; the light red and green parts around the N atoms represent the accumulation of positive and negative charge, respectively. It is clear that the N atoms become polarized with the merging anharmonicity, and this is consistent with what we know from the animation of the atomic movement. Then, when we use a third-order polynomial function to fit the curve, the fit does not improve much. It is only when the fourth-order polynomial function is adopted that the energy-displacement curve is perfectly fitted. It is apparent that, in addition to the three-phonon processes, fourth-order anharmonic effects also play an important role in the integral anharmonicity of TiNBr.

\section{Conclusions}

In summary, we have performed a comprehensive study on the thermoelectric properties of TiNBr including the band structures, the phonon dispersion, the mode Gr\"{u}neisen parameters, and other thermoelectric properties using DFT and solving the Boltzmann transport equation with first-principles calculations. TiNBr tends to form p-type semiconductors with a band gap of 1.299 eV. The absolute value of the Seebeck coefficient near the Fermi level at 300K can be as high as 2215 $\mu V/K$, which corresponds to n-type doping. The electronic structure of TiNBr indicates that the electrical properties can be enhanced by intercalation between the two adjacent bromide layers.

\par

Phonon dispersion and the corresponding PDOS indicate the thermodynamic stability of TiNBr. By solving the phonon BTE, we get low values of 1.34 W/mK, 1.59 W/mK, and 0.49 W/mK at 800K for the a, b, and c axes, respectively, for the intrinsic lattice thermal conductivity. Combined with the former electrical properties, these yield promising values of 0.66, 0.43, and 0.42 for the figure of merit (ZT) along the a, b, and c axes, respectively, at 800K if we conservatively set the electronic relaxation time as 0.8$\times 10^{-14}$s.

\par

To understand the underlying mechanism for the low $\kappa_l$ of TiNBr, a systematic mode level analysis based on the phonon group velocity, Gr\"{u}neisen parameter, phonon scattering rate, and vibrational modes is performed. The root cause of the low $\kappa_l$ for TiNBr is the collectively low phonon group velocity ($2.05\times 10^3 $ m/s on average) and the large phonon anharmonicity, which can be quantified using the Gr\"{u}neisen parameter and three-phonon processes. The animation of the atomic motion in highly anharmonic modes mainly involves the motion of the N atoms, and the charge density difference reveals that the N atoms become polarized with the merging of anharmonicity. Moreover, the fitting procedure for the energy-displacement curve verifies that, in addition to the three-phonon processes, the fourth-order anharmonic effect is also important in the integral anharmonicity of TiNBr.

\par

The significance of this paper is that, for the first time, we conducted comprehensive and systematic first-principles calculations on a material for which the thermoelectric properties have never been investigated. Furthermore, we have established a connection between the low lattice thermal conductivity and the behavior of phonon vibrational modes. The promising high Seebeck coefficient, the ZT value, and low thermal conductivity of TiNBr will benefit applications of this material in the field of energy conversion.

\begin{acknowledgments}
The authors thank Dr. Yandong Sun and Dr. Kerong Hu  of Tsinghua University for their helpful discussions.
This work was supported by the National Key Research Programme of China, under grant No. 2016YFA0201003,
Ministry of Sci \& Tech of China through a 973-Project under grant No. 2013CB632506, NSF of China
(51672155 and 51532003) and by the Tsinghua National Laboratory for
Information Science and Technology. The calculations were also conducted at the National Supercomputer
Center in Tianjin, and the calculations were performed on TianHe-1(A).
\end{acknowledgments}

\newpage 

\end{document}